\RequirePackage{fix-cm}
\documentclass[smallextended]{svjour3}       % onecolumn (second format)
\smartqed  % flush right qed marks, e.g. at end of proof
\usepackage{geometry,amsmath}
\usepackage{amssymb}
\usepackage{booktabs}
\usepackage{lineno}
\usepackage{graphicx}
\usepackage{multirow}
\usepackage{multicol}
\usepackage{setspace}
\usepackage{enumerate}
\usepackage{color}
\usepackage{algorithm, algorithmic}
\newcommand{\my}{\hangafter 1
\hangindent 1.6em
\noindent}
\journalname{Neural Processing Letters}
\begin{document}

\title{An Ensemble Classification Algorithm Based on Information Entropy for Data Streams%\thanks{Grants or other notes
}


\author{Junhong Wang         \and
        Shuliang Xu          \and
        Bingqian Duan        \and
        Caifeng Liu          \and
        Jiye Liang           \and
}

\institute{J. Wang \at
              1. School of Computer and Information Technology, Shanxi University, Taiyuan, China \\
              2. Key Laboratory of Computational Intelligence and Chinese Information Processing, Ministry of Education, Taiyuan, China\\
              Tel.: +86 0351-7010566\\
              \email{wjhwjh@sxu.edu.cn}           %  \\
           \and
           S. Xu \at
              School of Computer and Information Technology, Shanxi University, Taiyuan, China\\
              \email{xushulianghao@126.com}
           \and
           B. Duan \at
               School of Computer and Information Technology, Shanxi University, Taiyuan, China\\
               \email{6206486@qq.com}
           \and
           C. Liu \at
               School of Computer Science and Technology, Faculty of Electronic Information and Electrical Engineering, Dalian University of Technology, Dalian, China\\
               \email{liucaifeng12345@qq.com}
           \and
           J. Liang \at
               Key Laboratory of Computational Intelligence and Chinese Information Processing, Ministry of Education, Taiyuan, China\\
               \email{ljy@sxu.edu.cn}
}

\date{Received: 02 Aug 2017 / Accepted: date}

\maketitle

\begin{abstract}
Data stream mining problem has caused widely concerns in the area of machine learning and data mining. In some recent studies, ensemble classification has been widely used in concept drift detection, however, most of them regard classification accuracy as a criterion for judging whether concept drift happening or not. Information entropy is an important and effective method for measuring uncertainty. Based on the information entropy theory, a new algorithm using information entropy to evaluate a classification result is developed. It uses ensemble classification techniques, and the weight of each classifier is decided through the entropy of the result produced by an ensemble classifiers system. When the concept in data streams changing, the classifiers' weight below a threshold value will be abandoned to adapt to a new concept in one time. In the experimental analysis section, six databases and four proposed algorithms are executed. The results show that the proposed method can not only handle concept drift effectively, but also have a better classification accuracy and time performance than the contrastive algorithms.
\end{abstract}

\section{Introduction}
%\linenumbers
With the development of information society, many fields have produced a large amount of data streams, such as: e-commerce, network monitoring, telecommunication, stock trading, etc. The data in these fields are very different from the conventional static data, due to achieving fast, unlimited number and concept drift in data streams, it makes the technologies of traditional data mining face a big challenge [1].

Since data streams mining problem was proposed, it has received many attentions [2-6]. Especially, it has been widely used when the ensemble classification method was put forward [7]. Li et al. proposed a data stream classification algorithm based on a random decision tree model [8]. The algorithm creates a number of random decision trees, then randomly chooses the split attribute, and determines the split value through the information gain; in addition, the algorithm can effectively distinguish noise data and concept drift. Elwell et al. proposed an incremental learning algorithm for periodic concept drift called Learn++.NSE [9]. Learn++.NSE preserves the historical classification models; when historical classifier to classify the data correctly, the classifier will be improved the weight; after classifying correctly many times, the weight of the historical classifier will reach the activation threshold which is predetermined by a user; the classifier is activated from sleeping state, and then joins to the system to participate in deciding the labels of the unlabeled data. Aiming at the problem of the unbalanced data stream classification, Rushing et al. proposed an incremental algorithm based on the overlaying mechanism called CBEA [10]. While the algorithm training a classifier, the training examples are also saved. When the number of classifiers reaches the threshold, the two most similar coverage sets are selected, and the oldest classifier will be deleted. The final classification results are determined by KNN algorithm. Gomes et al. proposed an adaptive algorithm based on the social network called SAE [11], the algorithm introduces the related concepts of social network and each sub classifier seeing as a node on the network, so a network consists of multiple classifiers. SAE algorithm sets a series of parameters to measure the performances of the classifiers, and updates the classifiers to adapt to the new concept. Brzezinski et al. proposed a data stream classification algorithm based on online data called OAUE [12], the algorithm uses mean square error to determine the weights of classification models; when a period of time for the detection coming, the replacement strategy is used to deal with concept drift. Farid et al. proposed an ensemble classification algorithm based on instances weighting mechanism [13], in order to detect outliers, the algorithm combines with the clustering algorithm, if a data point does not belong to any existed cluster in the system, the class of the data will be thought as a new concept, and then the algorithm counts the data information in leaf nodes to further confirm the result. Liang et al proposed an online sequential learning algorithm called OS-ELM [14-17], OS-ELM is a development of the extreme learning machine (ELM) algorithm [18-22]; when a new data block coming, OS-ELM uses new data to incrementally update the structures of the single hidden feedforward neural networks; by the mean of the online sequential learning mechanism, OS-ELM can effectively deal with concept drift in data stream environment.

For the above ensemble classification algorithms, most of them detect the concept drift based on the accuracy of the classification result. In traditional data stream classification algorithm, accuracy is an important indicator which is used to detect the concept drift, but by using accuracy, it can only characterize current performance, and can not reflect the amount of information contained in the information system. Information entropy can be used to measure the degree of uncertainty of a system, and it has been proved to be an effective method to describe the information content in information systems [38-41]. Information entropy is a powerful tool to deal with uncertainty problem, and it has been applied in many fields. Aiming at the problem of the data stream classification, in this paper, we extended information entropy to measure the uncertainty of the concept in data stream, and then an ensemble classification algorithm based on information entropy (ECBE) is proposed. The new algorithm is based on ensemble classification technique; the weighted voting rule is adopted, and the weights of classifiers are determined according to the change of entropy values before and after classification; by using Hoeffding bound, ECBE algorithm can estimate whether concept drift appears or not. When concept drift appearing, the algorithm automatically adjusts the classifiers according to their weights. Comparing with the existed algorithms, the ECBE algorithm not only can effectively detect the concept drift, but also can get better classification results.

The rest of the paper is organized as follows: in section 2, we describe the related backgrounds; section 3 introduces the ensemble classification algorithm based on entropy for data stream; section 4 is the experimental process and data analysis; finally, section 5 gives the conclusions and summary.

\section{Backgrounds}
\subsection{Data stream and concept drift}
We assume $\left \{ \cdots,d_{t-1},d_{t},d_{t+1},\cdots  \right \}$ is a data stream generated by a system, where $d_{t}$ is the instance generating at \emph{t} moment; for every instance, we mark $d_{t}=\left \{ x_{t},y_{t} \right \}$, where $x_{t}$ is the features vector, and $y_{t}$ is the label. In order to illustrate related notions, we have the following definitions.

\textbf{Definition 1.} A certain number of instances are organized as a data set according to the time sequence, so we call the data set as data block, which denoted by $\left \{ d_{1},d_{2},\cdots,d_{n}\right \}$, where \emph{n} is the size of data block.

In the data stream classification, because in the face of massive data, the storage space of the data is far beyond the capacity of the computer memory, in order to make the algorithm handle massive data, the sliding window mechanism is widely used [23], in other words, each time, the window is only allowed one data block coming into the system, only if the data blocks in the current window are processed completely, the next data blocks can be acquired.

\textbf{Definition 2.} At \emph{t} moment, the data in sliding window is used to train a classifier, and the target concept we get is \emph{M}; $\Delta t$ time later, we use new data to train another classifier, and we get the target concept is \emph{N}; if $M\neq N$, we can say concept drift has happened in data stream. According to the different of $\Delta t$, concept drift can be divided into two types: when $\Delta t$ is a short time, concept drift is called gradual concept drift; when $\Delta t$ is a long time, concept drift is called abrupt concept drift [24,42] \footnote{http://kns.cnki.net/KCMS/detail/detail.aspx?dbcode=CDFD\&dbname=CDFD1214\&filename=1013198451\\.nh\&uid=WEEvREcwSlJHSldRa1FhcEE0NXdoZ2VUa09aRHFDZ2h2cVlMcHpDU3FSYz0=\$9A4hF\_YAuvQ5ob\\gVAqNKPCYcEjKensW4ggI8Fm4gTkoUKaID8j8gFw!!\&v=MjE2MDRIYkt4RnRYSnJwRWJQSVI4ZVgxTHV4W\\VM3RGgxVDNxVHJXTTFGckNVUkwyZllPUnRGeW5tVTczQVZGMjY=}

After concept drift appearing, the distribution of data in sliding window has changed, and $p_{t}\left ( y\mid x \right )\neq p_{t+1}\left ( y\mid x \right )$; at this time, the performance of the classifiers will decline, if the corresponding measures do not be taken, the error rate of the classification results will continuously rise. Therefore, in many applications, this property is used to detect concept drift [25].

\subsection{Ensemble classification techniques for data streams}
Ensemble classification is an important classification method for data streams, which uses a number of weak classifiers to combine into a strong classifier; so the method can effectively deal with the problem of concept drift. In the process of classification, the ensemble classification gives different classifiers with different weights, by adjusting the weights of the sub classifiers to update classifiers system to adapt to the new concept of a data stream, classification result of the data eventually is decided by voting mechanism [26]. In the data stream environment, the classification performance of ensemble classifiers is better than that of single classifier [27]. Because of the many advantages of ensemble classification, the method is widely used in data stream data mining [28-30].

\section{Ensemble classification based on information entropy in data streams}
\subsection{Concept drift detection based on information entropy}

Entropy is used to describe the disordered state of thermodynamic system. In physics, entropy is used to indicate the disorder of a system. Shannon extended the concept of the entropy to the information system, and utilized information entropy to represent the uncertainty of a source [31]. The entropy of a variable is greater, and it indicates that the uncertainty of this variable is larger, so it needs more information when changing the state of the variable from uncertain to certain.

In the classification of data streams, when data blocks are in the sliding window, at \emph{t} moment, we assumes a data block in the sliding window is $B_{i}$, and $B_{i}=\left \{ (\textbf{\emph{x}}_{i1},y_{i1}),(\textbf{\emph{x}}_{i2},y_{i2}),\cdots ,(\textbf{\emph{x}}_{in},y_{in}) \right \}$, where \textbf{\emph{x}} is the feature matrix of the ith data block, $\textbf{\emph{x}}=\left \{\textbf{\emph{x}}_{i1};\textbf{\emph{x}}_{i2};\cdots ;\textbf{\emph{x}}_{in} \right \}$, and \textbf{\emph{y}} is a label set of the ith data block, $\left \{ y_{i1}, y_{i2},\cdots, y_{in} \right \}\subseteq \textbf{\emph{y}}$. So at this time, the entropy of the data in the sliding window is calculated as the equation (1):
\normalsize
\begin{equation}
  H=-\sum_{j=1}^{\left | y \right |}p_{j}\log p_{j}
\end{equation}

Where $\left | \textbf{\emph{y}} \right |$ is the number of the labels in the $\textbf{\emph{y}}$, and $p_{k}$ is the probability of the label $y_{k}$ in the instances; $p_{k}$ can be calculated as the equation (2):

\begin{equation}
  p_{j}=\frac{\sum_{m=1}^{n}\left | y_{im}=y_{ij} \right |}{n}
\end{equation}

In the equation (2), if $y_{im}=y_{j}$, $\left | y_{im}=y_{ij} \right |=1$; else if $y_{im}\neq y_{j}$, $\left | y_{im}=y_{ij} \right |=0$.

By utilizing the classifiers to classify the data block, the classification result can be obtained with a weighted voting mechanism; at the same time, we can use the equations (1)-(2) to calculate the entropy values of the classification result and the real result of the data block which are marked as $H_{1}$ and $H_{2}$ respectively. The entropy's deviation of the data block can be calculated:
\begin{equation}
  \Psi (H)=\left | H_{1}-H_{2} \right |
\end{equation}

\textbf{Lemma} (Hoeffding bound) [32, 33]. By independently observing a random variable \emph{r} for \emph{n} times, the range of the random variable \emph{r} is \emph{R}, and the observed average of \emph{r} is $\bar{r}$; when confidence level is $1-\alpha$, the true value of \emph{r} is at least $\bar{r}-\varepsilon$, where $\varepsilon =\sqrt{\frac{R^{2}\ln (\frac{1}{\alpha })}{2n}}$.

\textbf{Theorem 1.} When the concept of a data stream (the distribution of the data) is stable, $S_{1}$ and $S_{2}$ are the deviation of the two adjacent data blocks which are calculated according to the equation (3), so it can be known that:
\normalsize
\begin{equation}
  \left | S_{1}-S_{2} \right |\leq 2\sqrt{\frac{R^{2}\ln (\frac{1}{\alpha })}{2n}}
\end{equation}

\textbf{Proof:} When the concept of a data stream is stable, the data distributions of the two adjacent data blocks are consistent, and the difference of the observed entropy values is very small. Let $S_{0}$ be the real entropy of the data block according to the equation (3), from the Hoeffding bound, it is known that:
\begin{equation*}
  \left | S_{1}-S_{0} \right |\leq \varepsilon \ \ \text{and} \ \ \left | S_{2}-S_{0} \right |\leq \varepsilon
\end{equation*}

So it have:
\begin{equation*}
  -\varepsilon \leq S_{1}-S_{0}\leq \varepsilon; \ \ -\varepsilon \leq S_{0}-S_{2}\leq \varepsilon
\end{equation*}

If plus the two inequalities, the result is as follow:
\begin{equation*}
  -2\varepsilon \leq S_{1}-S_{2}\leq 2\varepsilon
\end{equation*}

So it can get $\left | S_{1}-S_{2} \right |\leq 2\varepsilon$.
\vspace{3ex}

Therefore, it can use $\Delta H=\left | \Psi (H_{1})-\Psi (H_{2}) \right | $ as a measure to detect the concept drift in data streams, where $\Psi (H_{1})$ and $\Psi (H_{2})$ are the deviation of the two adjacent data blocks. If $\Delta H> 2\varepsilon$, it is thought concept drift has appeared. But it is obvious that when a special situation occurring, only using $\Delta H$ cannot work well: if the two adjacent data blocks represent different concepts, so the classification accuracies of the two adjacent data blocks are very low and they are almost equal; while calculating $\Delta H$, it is still less than or equal to $2\varepsilon$, and the algorithm will think no change happening, but in fact, concept drift has appeared at this time. In order to solve the problem, we correct the decision criterion of concept drift:
\begin{equation}
 \Delta H > 2\varepsilon ~\text{or}~\Psi (H_{1})> 2\varepsilon~\text{or}~\Psi(H_{2})> 2\varepsilon
\end{equation}

\noindent we can say concept drift has appeared. From the equation (5), it is known that when the two adjacent data blocks occuring two concept drifts, $\Delta H$ will be less than or equal to $2\varepsilon$ , however, $\Psi (H_{1})$ or $\Psi(H_{2})$ will be still greater than $2\varepsilon$, so concept drift will be detected.

For each sub classifier \emph{ensemble(j)}, $(j=1,2,3,\cdots,k)$; according to the classification results of the current data block, the entropy of the classification result and real result marked as $h_{i}$ and $H_{i}$ can be calculated by using the equation (1); so the change of the entropy before and after classification is $\Psi (h_{i})=\left | h_{i}-H_{i} \right|$, and the weights of the classifiers are updated according to the equation (6):
\normalsize
\begin{equation}
  ensemble(j).weight(t)=\delta (\Psi (h_{i}))\cdot ensemble(j).weight(t-1)
\end{equation}

In the equation (6), $ensemble(j).weight(t)$ is the new weight, and $ensemble(j).weight(t-1)$ is the weight before updating. The function $\delta (\Psi (h_{i}))$ is showed as the equation (7) :

\begin{equation}
\delta (\Psi (h_{i}))=
\left\{
\begin{aligned}
 &\frac{1}{e^{1+\Psi (h_{i})}} \ \ \ \ \ \ \ \ \ \ \ \ \ \ \ \ \ \ \ \ \ \ \ \ \ \ \ \ \ \Psi(h_{i})> 0\\
 &\beta,\ \ \text{$\beta$ is constant and $\beta >1 \ \ \ \ \Psi (h_{i})= 0$}
\end{aligned}
\right.
\end{equation}

\textbf{Theorem 2.} If the concept of the data stream is stable, $w_{1},w_{2},w_{3},\cdots ,w_{n}$ are the weights of the ensemble classifiers which are updated on the basis of the equation (6), and the confidence level is $1-\alpha$, for each sub classifier, the weight must satisfy the following condition:
\begin{equation*}
  w_{i}\geq \frac{1}{n}\sum_{q=1}^{n}w_{q}-\frac{S}{\sqrt{n}}t_{\alpha }(n-1)-3S
\end{equation*}

\begin{equation*}
  S^{2}=\frac{1}{n-1}\sum_{m=1}^{n}(\frac{1}{n}\sum_{q=1}^{m}w_{q}-w_{m})^{2}.
\end{equation*}
\textbf{Proof: }If the concept of a data stream is stable, each sub classifier in the system is adaptive to the current concept; therefore, the difference of the classifier¡¯ weights before after classifying is very small. We assume that the weights of the ensemble classifiers according with a normal distribution, so the mean and variance of the weights are calculated as:

\begin{equation*}
  \mu =\frac{1}{n}\sum_{q=1}^{n}w_{q} \ \ \text{and} \ \ S^{2}=\frac{1}{n-1}\sum_{m=1}^{n}(\mu -w_{m})^{2}
\end{equation*}

So we have:
\begin{equation*}
  P\left \{ \frac{\mu -w}{\frac{S}{\sqrt{n}}} \leqslant t_{\alpha}(n-1) \right \}=1-\alpha
\end{equation*}

From the inequality, it can be known
\begin{equation*}
  P\left \{w\geq \mu-\frac{S}{\sqrt{n}}  t_{\alpha}(n-1) \right \}=1-\alpha
\end{equation*}

At the level of confidence $1-\alpha$, we have

\begin{equation*}
  w \geq \mu -\frac{S}{\sqrt{n}}t_{\alpha }(n-1)
\end{equation*}

We use \emph{S} to approximately replace the standard deviation of the normal distribution $\delta$, according to the $3\delta$  principle of the normal distribution, it can conclude:

\normalsize
\begin{equation*}
  w_{i}\geq w-3S\geq \frac{1}{n}\sum_{q=1}^{n}w_{q}-\frac{S}{\sqrt{n}}t_{\alpha }(n-1)-3S.
  \vspace{3ex}
\end{equation*}

From the theorem 2, we can see that when concept drift appearing, the algorithm will obtain the statistic about the weights of all classifiers from this time to the last time when the last concept drift happening, and the lower limit of the updating weight for each sub classifier is calculated as follow:

\begin{equation}
  \theta\_weight=\frac{1}{n}\sum_{q=1}^{n}w_{q}-\frac{S}{\sqrt{n}}t_{\alpha }(n-1)-3S
\end{equation}

According to the classification result of sub classifiers for the new data, the weights are updated by using the equation (6). When the classifiers cannot adapt to the current concept, the weights of the classifiers will sharply decreases below $\theta\_weight$. Finally, the system will delete all classifiers which does not satisfy the equation (8), so in the next process of the classification, the classifiers with low performances do not continue to participate in the decision-making.

\subsection{The execution of the ECBE algorithm}
From the above knowledge, we can know that the implementation steps of ECBE are as follows:
\begin{algorithm}
	\renewcommand{\algorithmicrequire}{\textbf{Input:}}
	\renewcommand{\algorithmicensure}{\textbf{Output:}}
	\caption{ECBE}
	\label{alg:1}
	\begin{algorithmic}[1]
    \REQUIRE Ensemble classifier: \emph{ensemble}=NULL; data stream \textbf{\emph{S}}; the number of sub classifier: \emph{k}; the size of data block: \emph{winsize}; the array preserving the weights of the classifiers in the period of two adjacent concept drifts: \emph{num}=NULL;
    \ENSURE The trained classifiers: \emph{ensemble}.
    \WHILE{\textbf{\emph{S}}!=NULL}
         \STATE Read winsize instances to organize a data block;
         \IF{size(\emph{ensemble})$<$ \emph{k}}
             \STATE Use $B_{i}$ to train a new classifier $C_{j}$;
             \STATE $ensemble\leftarrow ensemble \ \cup \ C_{j}$ and $C_{j}.weight=1$;
         \ELSE
             \STATE Use the equation (3) to calculate the entropies $\Psi (H_{1})$ and $\Psi (H_{2})$;
             \FOR{$ \textbf{each} \ \ ensemble(t) \in ensemble$  //updating the weights of the classifiers}
                 \STATE  Calculate the entropy $h_{t}$ of the result which is from the \emph{ensemble(t)} classifying the data block $B_{i}$, and update the weights of the classifiers according to the equation (6);
                 \STATE $num\leftarrow num \cup ensemble(t).weight$;
            \ENDFOR
            \IF{$\Delta H > 2\varepsilon ~\text{or}~\Psi (H_{1})> 2\varepsilon~\text{or}~\Psi(H_{2})> 2\varepsilon$  // concept drift}
               \STATE Delete the classifier with the minimum weight;
               \STATE Calculate the weight $\theta \_weight$ according to the equation (8);
               \STATE num=NULL;
            \ENDIF
            \STATE Delete the classifier with weights less than $\theta \_weight$;
            \STATE Use $B_{i}$ to train a new classifier $C_{new}$;
            \IF{size(\emph{ensemble})$>=$\emph{k}}
               \STATE  Delete the classifier with the minimum weight;
            \ENDIF
            \STATE $ensemble=ensemble\cup C_{new};$
            \STATE Use $B_{i}$ to train each classifier in the system;
       \ENDIF
    \ENDWHILE
    \end{algorithmic}
\end{algorithm}

In the ECBE algorithm, it uses entropy to detect concept drift; if the data block does not appear concept drift, the classifiers system will have a good performance, so the classification results and the actual labels are nearly consistent, and the difference of the two entropies is very small. When concept drift appearing, the error rate of the classification result will increase; the difference of the entropy will also increase, the increment will promote that concept drift in the data stream can be detected. After concept drift is detected, some classifiers in the system are unable to adapt to the current concept; if the defunct classifiers are not eliminated, it will not only decreases the accuracy of the classification results, but also gives a false alarm about concept drift. In order to solve this problem, ECBE algorithm preserves all the weights of the classifiers where the concept of the data is stable in a period of two adjacent concept drift. When concept drift happens, by saving the statistical weights, the lower bound of the weight when the concept is stable can be computed. Meanwhile, because of the concept changing, the weights of the classifiers will be dramatically reduced, so all the classifiers whose weights are below the lower bound will be deleted; the measure ensures that the classifier system can adapt to new concept at a very fast speed.

\section{Experimental analysis}

In order to validate the performance of ECBE algorithm, in this paper, we chosen SEA [34], AddExp [35], AWE [27], DCO [36],WEAP\_I[4], TOS-ELM [20] and OS-ELM [14] as the comparison algorithms; C4.5, CART and Naive Bayesian et al. are used as base classifiers; all the algorithms were tested on the three artificial datasets: \emph{waveform}, \emph{Hyperplane} and \emph{LED} which were produced by MOA platform [37] and the three practical datasets: \emph{sensor\_reading\_24}, \emph{shuttle} and \emph{page-blocks}. The parameters of the experimental environment are as follows: Windows7 operating system, Intel dual core CPU, 4 G memory. The algorithm is implemented by R2013a Matlab.

\subsection{The descriptions of the datasets}

The datasets using in the experiment are cited from UCI datasets. We noly give a brief explain about the datasets, and the detail can be see from the website \footnote{http://archive.ics.uci.edu/ml/datasets.html}.

\emph{waveform} dataset: it is an artificial dataset and the data set has 50000 instances. Each instance has 22 attributes; the first 21 attributes are numeric and there are 3 different labels in the dataset.

\emph{LED} dataset: the dataset has 50000 instances, each of which contains 25 attributes, and the values of the first 24 attributes are 0 or 1. There are 10 different labels in the dataset; the data contains 5\% of the noise.

\emph{Hyperplane} dataset: a sample X in a d dimensional hyperplane satisfies the following mathematical expression: $\sum_{i=1}^{d}a_{i}x_{i}=a_{0}$, where $a_{0}=\frac{1}{2} \sum\nolimits_{i=1}^{d}a_{i}$. The dataset contains 50000 instances and 11 dimensions; the values   of the first 10 dimensions are in [0, 1]. If $\sum_{i=1}^{d}a_{i}x_{i}\geq a_{0}$, the label of the instance is marked as positive, otherwise, the label of the instance is marked as negative. In addition, the dataset contains 10\% noise.

\emph{sensor\_reading\_24} dataset: the dataset contains 5456 instances and 25 properties. The first 24 properties are real; there are 4 different labels in the dataset.

\emph{shuttle} dataset: the dataset contains 43500 instances and 11 attributes. The values of the first 10 attributes are continuous; there are 7 different labels in the dataset.

\emph{page-blocks} dataset: the dataset contains 5473 instances and 11 attributes. In the first 10 attributes, some attributes values are continuous, and some of them are discrete. There are 5 different labels in the dataset.

\subsection{The experimental results}

In order to verify the performance of the ECBE algorithm, SEA, AddExp, AWE, DCO, WEAP\_I, TOSELM and ECBE are run on the three artificial datasets: \emph{waveform}, \emph{Hyperplane} and \emph{LED}. For all the algorithms, \emph{k} = 5, \emph{winsize}=2000; the parameters of AddExp are as follows:  $\beta =0.5$, $\gamma=0.1$; the parameters of ECBE are as follows: $\alpha=0.05$, $\beta =0.5$, $\gamma=0.1$; the test results are showed in Tables 1 and 2.

\begin{table*}[!h]
\footnotesize
\centering
\caption{ Average accuracy on the artificial datasets}
\begin{tabular}{cccc}
  \bottomrule
  &waveform &Hyperplane &LED \\\midrule
  SEA    &0.6777    &0.6925 &0.2878 \\
  AddExp &0.3701    &0.5404 &0.1622 \\
  AWE	 &0.6653	&0.6912	&0.2715 \\
  DCO	 &0.6628	&0.7086	&0.3040 \\
  WEAP\_I&0.7090    &0.6349	&0.4165 \\
  TOSELM &0.5411	&0.6349	&0.3143 \\
  ECBE	 &\textbf{0.7583} &\textbf{0.7109}	&\textbf{0.8145} \\\bottomrule
\end{tabular}
\end{table*}

\begin{table*}[!h]
\footnotesize
\centering
\caption{Time overhead on the artificial datasets (Unit: sec)}
\begin{tabular}{llll}
  \bottomrule
  &waveform &Hyperplane &LED \\\midrule
  SEA &2743.422	&1429.004	    &6.2156 \\
  AddExp &1124.080	&1123.530	&1249.691\\
  AWE	&2272.473	&1004.433	&4.8481 \\
  DCO	&1084.533	&407.561	&2.5660 \\
  WEAP\_I&225.9743  &99.3225	&4.0223 \\
  TOSELM	 &\textbf{0.01633}	&\textbf{0.0338}	    &\textbf{0.1407} \\
  ECBE	&164.946	&85.381    &26.551 \\\bottomrule
\end{tabular}
\end{table*}

From Tables 1 and 2, it is known that, on the three artificial datasets, for ECBE, the accuracy is the highest and the time overhead is far less than the other compared algorithms on most datasets except for TOSELM; the comprehensive performance of ECBE is the best of all.

From the data Table 1, comparing with the other algorithms, AddExp has the lowest accuracy, and it is time consuming. The reason for the result is that AddExp updates the classifiers based on a single data; on the three datasets, the data contains gradual concept drift, and the concept of the data is constantly and slowly changing, which makes the accuracy of the old classifier decreasing; when an instance is classified wrongly by a classifier, the algorithm would eliminate the classifier to deal with the concept drift, however, the classifiers in the AddExp algorithm can get a better performance only after trained many times; due to the training for the new classifiers is inadequate, the error rate of the classification result is very higher, which will lead to the more classifier being replaced, so all the classifiers filling in the system are not trained adequately and the AddExp algorithm has a higher error rate on the three datasets.

From Table 1, for the \emph{LED} dataset, the accuracies of the algorithms are very low except the ECBE algorithm. On the \emph{LED} dataset, there are 7 attributes relating to concept drift, and the others are redundant attributes. When concept drift appearing, the speed of the concept changing on the \emph{LED} dataset is faster than that on the waveform and Hyperplane datasets; the three algorithms: SEA, AddExp and AWE delete the classifiers with low performance on the basis of a single classifier, so it will take a long training time to completely eliminate the weak classifiers; when the speed of concept drift is lower than the elimination speed, the performance of the classifiers will be always at a low level. For the ECBE algorithm, when a concept drift is detected, the weights of the classifier will be greatly reduced, however, ECBE will delete all weak classifiers according to their weights at a time, so ECBE can adapt to the new concept at a fast speed. In this experiment, it shows the performance of ECBE is the best of all.

In order to verify the effectiveness of the algorithm to deal with the practical datasets, SEA, AddExp, AWE, DCO, WEAP\_I, TOSELM and ECBE are run on the \emph{sensor\_reading\_24}, \emph{shuttle} and \emph{page-blocks} datasets, For all the algorithms, the maximum number of classifiers \emph{k}=5; in the \emph{sensor\_reading\_24} and \emph{page-blocks} datasets, the size of data block \emph{winsize}=200 and on the others datasets, \emph{winsize}=2000; the parameters of AddExp are as follows: $\beta=0.5$, $\gamma =0.1$; the parameters of ECBE are as follows: $\alpha=0.05$, $\beta=2$, $\gamma =0.1$; the obtained results are shown in Tables 3 and 4.

\begin{table*}[!h]
\footnotesize
\centering
\caption{Average accuracies on the practical datasets}
\begin{tabular}{cccc}
  \bottomrule
  &sensor\_reading\_24 &shuttle &page-blocks \\\midrule
  SEA &0.8168 &0.8717 &0.8923 \\
  AddExp &0.5715 &0.8127 &0.7877\\
  AWE	&\textbf{0.8673}	&0.8691	&\textbf{0.9235} \\
  DCO	&0.8458	&0.8615	 &0.9231 \\
  WEAP\_I&0.9500  &0.9054 &0.9035 \\
  TOSELM	 &0.5608 &0.8333 &0.8973 \\
  ECBE	&0.7939 &\textbf{0.9137}	&0.9215 \\\bottomrule
\end{tabular}
\end{table*}

\begin{table*}[!h]
\footnotesize
\centering
\caption{Average time overhead on the real datasets (Unit: sec)}
\begin{tabular}{cccc}
  \bottomrule
  &sensor\_reading\_24 &shuttle &page-blocks \\\midrule
  SEA &269.944	&1163.610	&98.826 \\
  AddExp &132.719	&1028.906	&127.921\\
  AWE	&235.112	&945.015	&66.186\\
  DCO	&78.760	&365.326	&20.060 \\
  WEAP\_I &9.7250  &54.9151  &3.9338 \\
  TOSELM  &2.1260  &2.1030  &1.2650 \\
  ECBE	&\textbf{1.678}	&\textbf{1.156}	&\textbf{1.013} \\\bottomrule
\end{tabular}
\end{table*}

From Tables 3 and 4, it can be known that all the algorithms get good performances on the practical datasets. The accuracy of ECBE is most prominent only on the \emph{shuttle} dataset, but every accuracy of the algorithm is more than 0.79 and the accuracies of ECBE on the other datasets are very close to the best results; in addition, for the time overhead, ECBE is far less than the comparison algorithms, so the advantage of ECBE is very obvious.

From the data in Tables 3 and 4, the performances of SEA, AddExp, AWE, DCO, WEAP\_I and TOSELM testing on the three practical datasets are very close and most of them have achieved good results. On the \emph{sensor\_reading\_24}, \emph{shuttle} and \emph{page-blocks} datasets, the data whose labels are the same is highly concentrated; in most cases, the data distribution of two adjacent data blocks is consistent, and the classification model trained by the data has a better adaptability for new data, so the most performances of the eight algorithms are very good on the practical datasets. After analyzing Tables 1-4, SEA, AddExp, AWE and DCO are effective for the datasets where the data with the same labels is highly concentrated, but when dealing with the datasets containing gradual concept drift, the performance of those algorithms are not satisfactory; if we analyze the results of ECBE, it is very obvious that ECBE can deal with the two types of concept drift.

In the ECBE algorithm, we use the equation (8) to calculate the threshold which judges the change of the entropy of data stream, and the threshold has an important influence on the performance of the algorithm; in the algorithm, the \emph{n} in the equation (8) is \emph{20 $\times$ winsize}. To explore the impact of the \emph{winsize} values for the algorithm, we select the \emph{waveform} dataset as the experimental data, and the parameters of ECBE are as follows: $\alpha=0.05$, $\beta=2$, $\gamma =0.1$; we run the ECBE algorithm, and the results are showed in Table 5 and Fig.1-6.

\begin{table*}[!h]
\footnotesize
\centering
\caption{Average accuracy with different \emph{winsize} values}
\begin{tabular}{cccccccccccc}
  \bottomrule
  winsize           &800	&1000	&1200	&1400	&1600	&1800 &2000 &2200 &2400\\\midrule
  Hyperplan &0.7050 &0.6968 &0.7051 &0.7081 &0.7041 &0.7132 &0.7109 &0.7065 &0.7080 \\
  LED       &0.5961 &0.8284 &0.7403 &0.8245 &0.8165 &0.8151 &0.8145 &0.8166 &0.8131 \\
  shuttle   &0.8922 &0.9143 &0.9002 &0.9013 &0.9115 &0.9042 &0.9979 &0.9145 &0.9604 \\
  waveform	&0.7602 &0.7565 &0.7559 &0.7556 &0.7561 &0.7561	&0.7583 &0.7519 &0.75\\\bottomrule
  winsize   &20	&40	&60	&80	&100 &120 &140 &160 &190\\\midrule
  page      &0.9409 &0.9220 &0.9275 &0.9303 &0.9096 &0.9150 &0.9109 &0.9287 &0.9193 \\
  sensor    &0.9474 &0.8000 &0.7349 &0.7913 &0.6758 &0.7724 &0.8083 &0.7664 &0.7623 \\\bottomrule
\end{tabular}
\end{table*}

\begin{figure*}[!h]
  \centering
  \includegraphics[width=8cm,height=6cm]{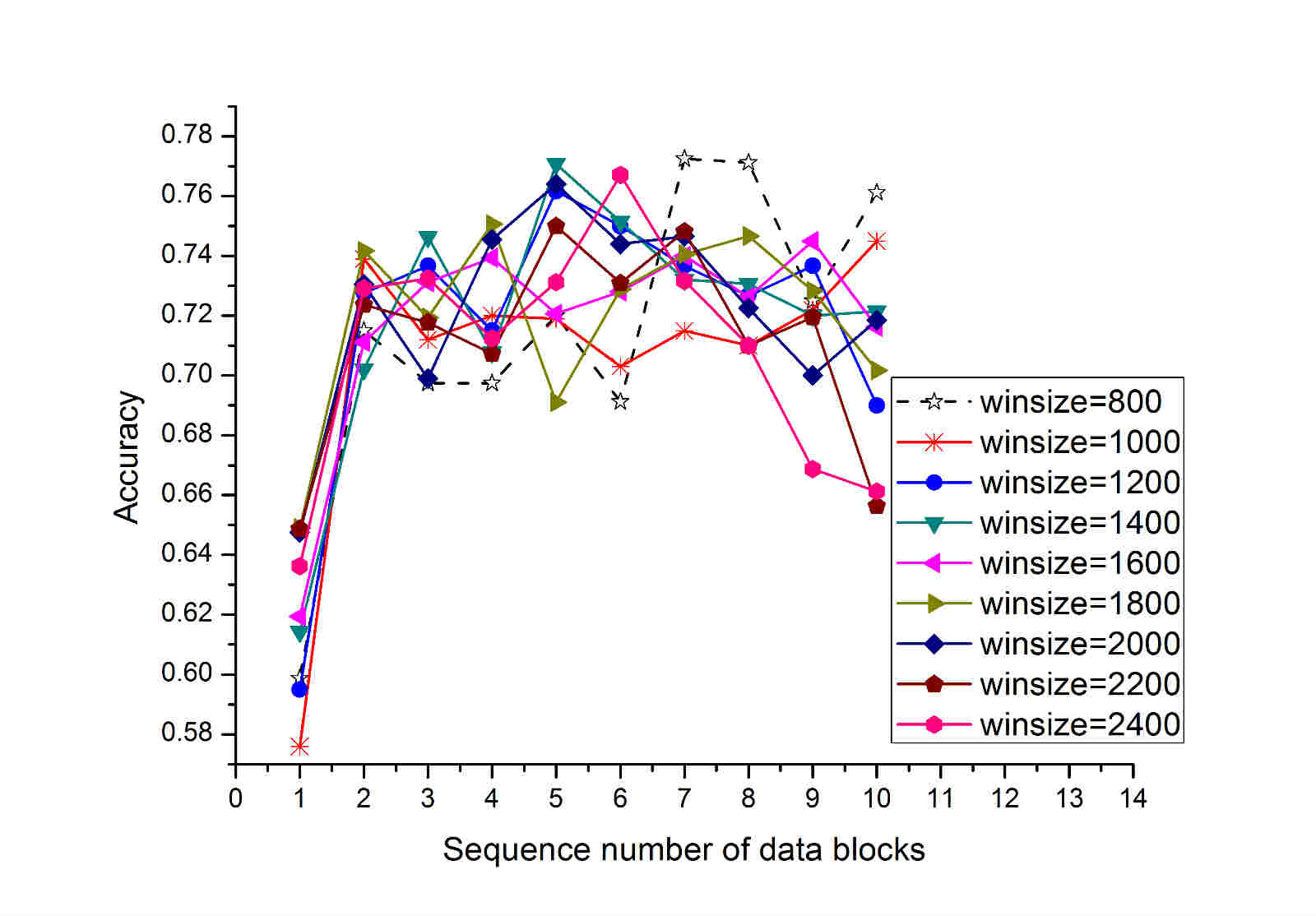}\\
  \caption{Test result with different \emph{winsize} values on Hyperplane}
\end{figure*}

\begin{figure*}[!h]
  \centering
  \includegraphics[width=8cm,height=6cm]{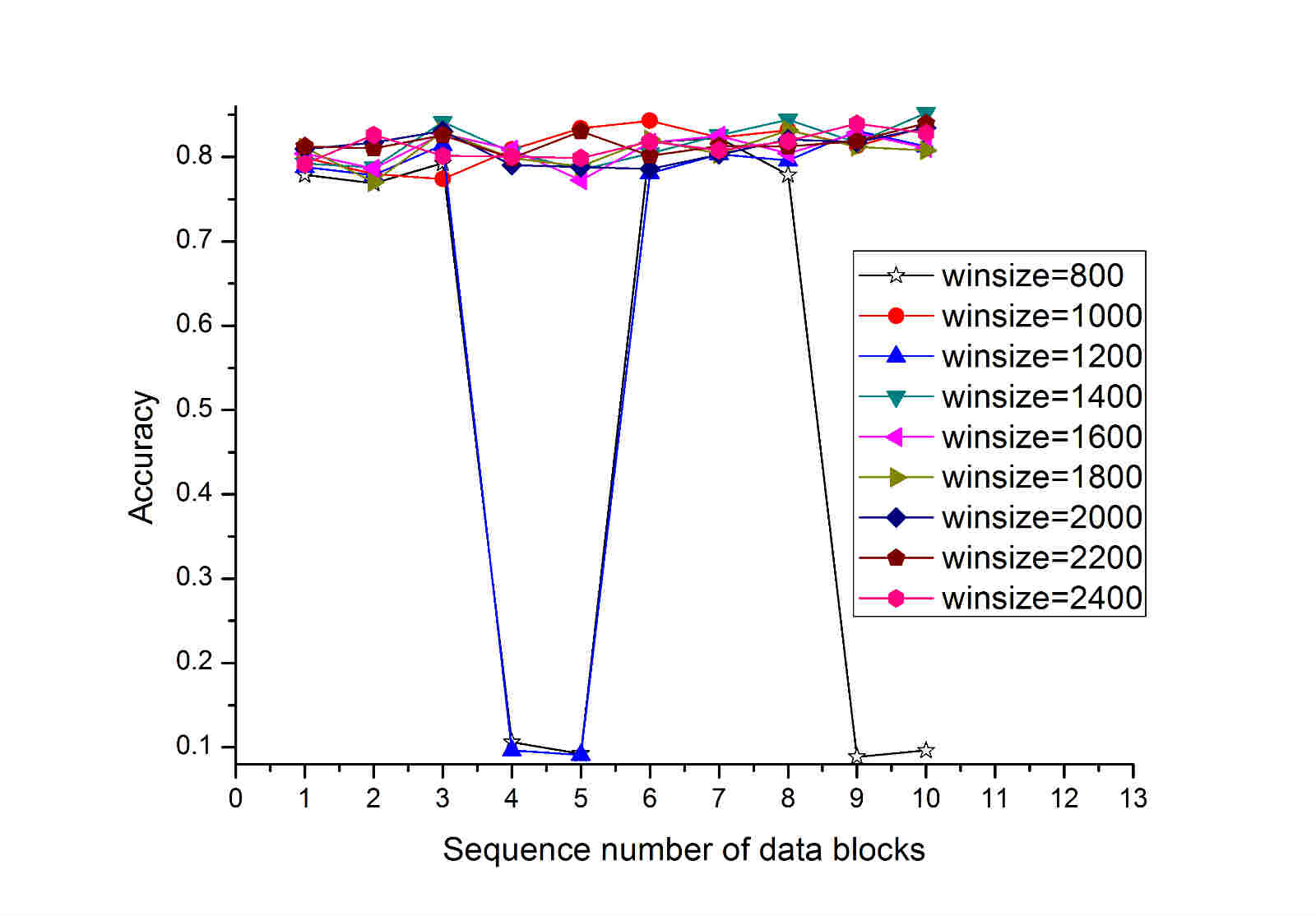}\\
  \caption{Test result with different \emph{winsize} values on LED}
\end{figure*}

\begin{figure*}[!h]
  \centering
  \includegraphics[width=8cm,height=6cm]{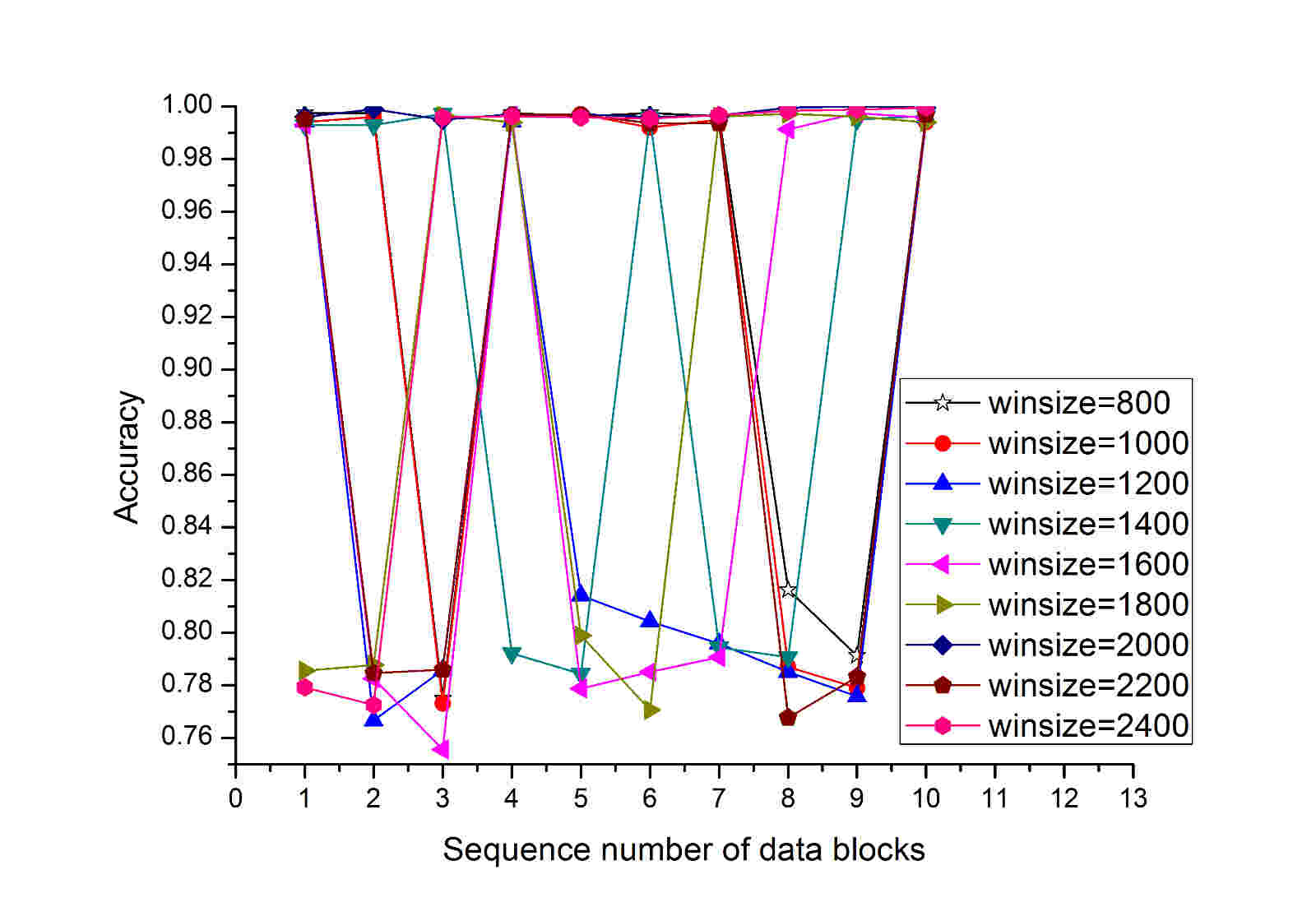}\\
  \caption{Test result with different \emph{winsize} values on shuttle}
\end{figure*}

\begin{figure*}[!h]
  \centering
  \includegraphics[width=8cm,height=6cm]{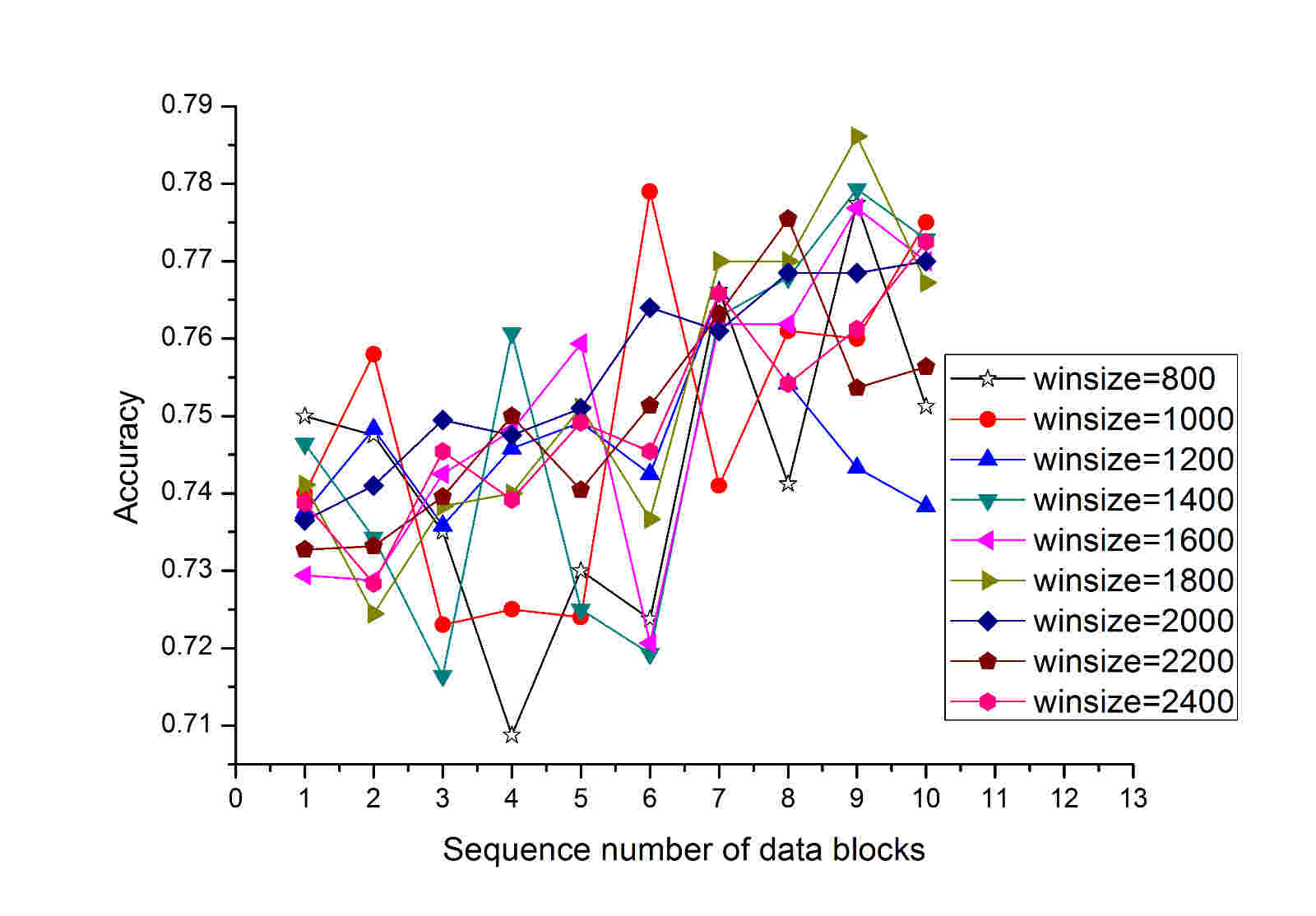}\\
  \caption{Test result with different \emph{winsize} values on waveform}
\end{figure*}

\begin{figure*}[!h]
  \centering
  \includegraphics[width=8cm,height=6cm]{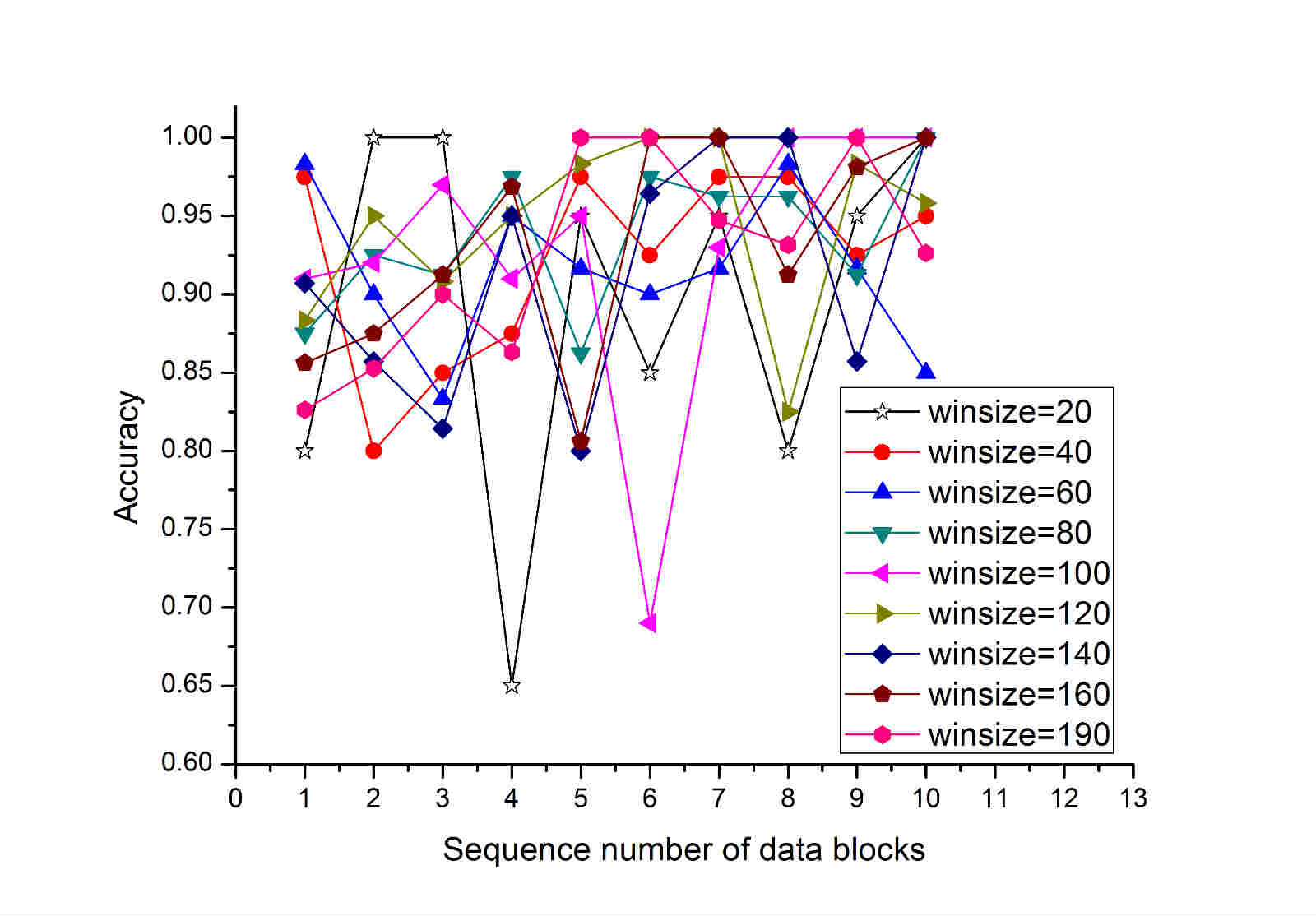}\\
  \caption{Test result with different \emph{winsize} values on page}
\end{figure*}

\begin{figure*}[!h]
  \centering
  \includegraphics[width=8cm,height=6cm]{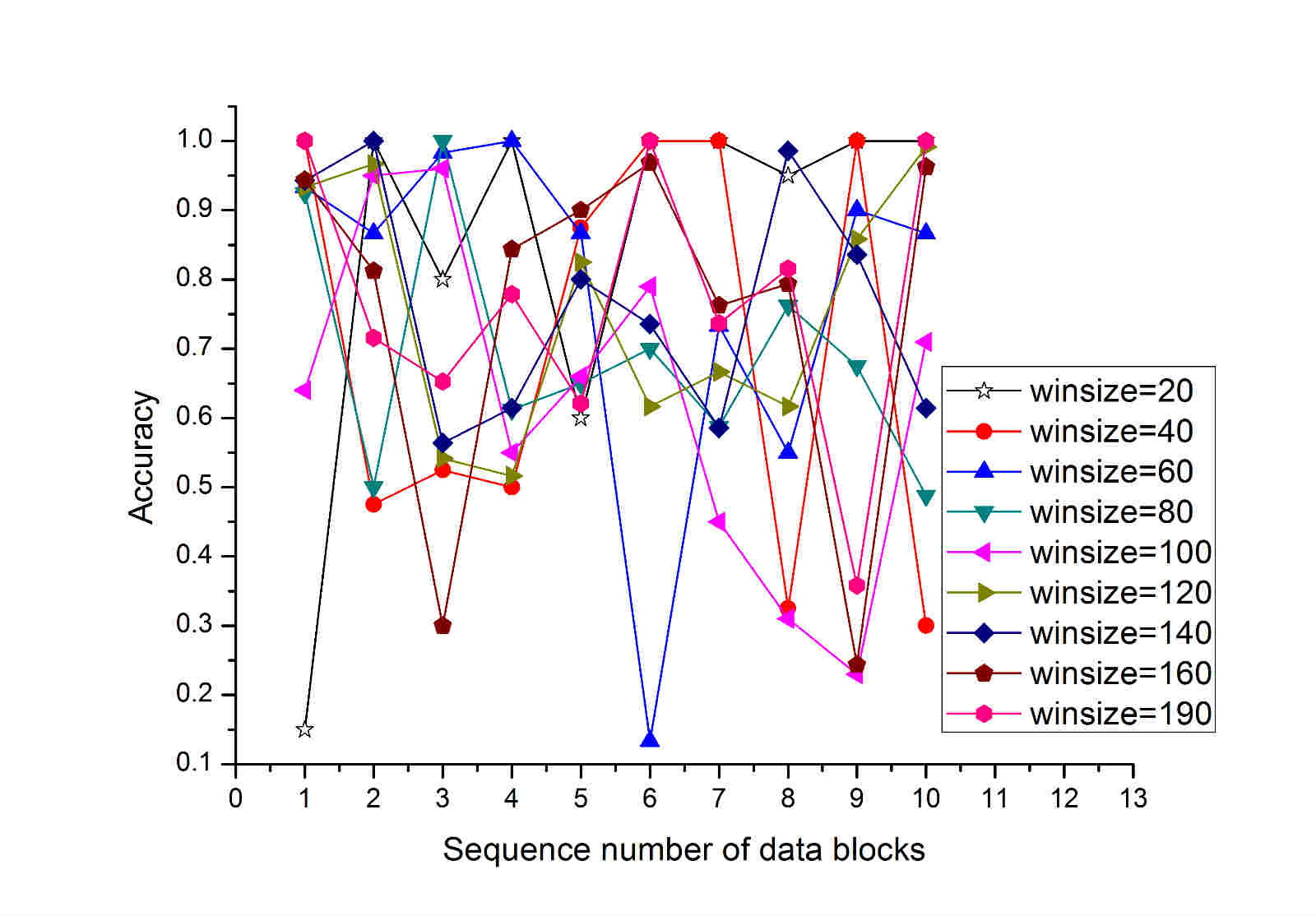}\\
  \caption{Test result with different \emph{winsize} values in Hyperplane on sensor\_reading\_24}
\end{figure*}

In the Fig.1-6 and Table 5, under the situation of different \emph{winsize} values, it can be seen, from the test result of the individual data blocks and the entire test results, there is no obvious linear rule between the winsize values and the performance of ECBE. For the ECBE algorithm, a small \emph{winsize} value can lead to a large threshold, which can make ECBE is not too sensitive to the change of data distribution, but a small data block will cause the classifiers trained by the data are under fitting, so the performance of the classifiers system will not be remarkably improved. If the \emph{winsize} value is very large, it will lead to a small threshold, which makes the algorithm is too sensitive to the change of data distribution, so the classifiers will give a false alarm of concept drift and affect the performance of ECBE. Therefore, on a practical application, the value of \emph{winsize} needs to consider the data distribution and the speed of concept changing; after trying many times, we select the optimal \emph{winsize} value from a series of candidate values.

In order to study the anti noise performance of ECBE, we select the \emph{waveform} dataset as the experimental data, and add 5\%, 10\%, 15\% and 20\% noise. The parameters of ECBE are set as follows: \emph{winsize}=2000, $\alpha=0.05$, $\beta=2$, $\gamma =0.1$. We run the ECBE algorithm on the datasets, and the test results are showed in the Fig.7 and Table 6.
\begin{figure*}[!h]
  \centering
  \includegraphics[width=8cm,height=6cm]{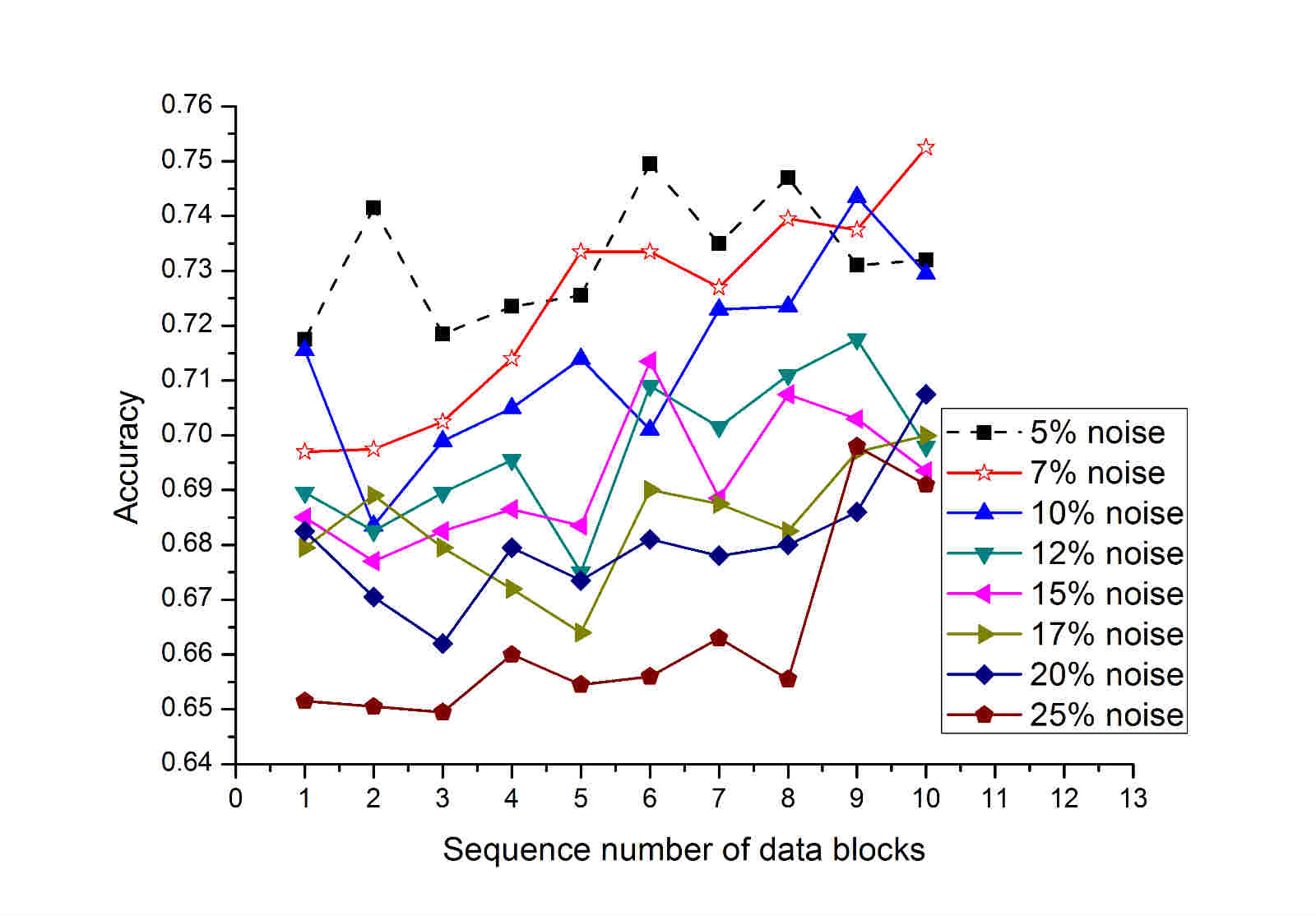}\\
  \caption{Test result with different noise level}
\end{figure*}

\begin{table*}[!h]
\footnotesize
\centering
\caption{Average accuracy of algorithm with different noise}
\begin{tabular}{cccccccccccc}
  \bottomrule
  Noise	            &5\%     &7\%	  &10\%	    &12\%    &15\%	  &17\%     &20\%     &25\%\\\midrule
  Average accuracy	&0.7348	 &0.7263  &0.7167	&0.7009  &0.6929  &0.6863   &0.6842   &0.6678\\\bottomrule
\end{tabular}
\end{table*}

From the Fig.7, it can be known that the accuracy of ECBE also gradually reduces with the increase of noise. From the table 6, we can see, comparing with the data containing 5\% noise, when the noise increasing 2\%, 5\%, 7\%, 10\%, 12\% ,15\% and 20\%, the accuracy of ECBE only decreases 0.0085, 0.0181, 0.0339,  0.0419, 0.0485, 0.0506 and 0.067 respectively and the magnitude of the change is very small, thus it is known that ECBE has a good noise immunity. The good noise immunity of ECBE comes from the elimination strategy; when the noise in the data stream is increasing, the difference of the entropies of the classification results and the actual results increases, and the weights of classifiers will decay at a fast speed. If the weights are below a specified threshold, the classifiers trained by the noise data will be deleted, so the weak classifiers will not affect the classification of new data.

In order to further test the performance of ECBE, we compare ECBE with a representative neural network algorithm called OS-ELM on the 6 datasets and test OS-ELM with different activation functions. In the experiment, we choose \emph{radbas, sigmoid, sine} and \emph{hardlim} functions as the activation functions. The parameters of ECBE are as follows:$\alpha=0.05$, $\beta=2$, $\gamma =0.1$; the test results are showed in Tables 7 and 8.

\begin{table*}[!h]
\footnotesize
\centering
\caption{ Average accuracy testing on datasets}
\begin{tabular}{cccccccccccccccccccc}
  \bottomrule
  \multicolumn{1}{c}{\multirow{2}*{}}  &\multicolumn{1}{c}{\multirow{2}*{ECBE}} &\multicolumn{4}{c}{\multirow{1}*{OS-ELM}} &\multicolumn{1}{c}{\multirow{2}*{winsize}}\\
  \cline{3-6}
  &&radbas &sigmoid &sine &hardlim\\\midrule
  waveform	&0.7505	&0.3270	&\textbf{0.7629}	&0.3595	&0.7116	&2500 \\
  Hyperplane	&0.7080	&0.7020	&\textbf{0.7274}	&0.6936	&0.6763	&2500\\
  LED	&\textbf{0.8119}	&0.2100	&0.7480	&0.4737	&0.4673	&2500\\
  sensor\_reading\_24	&\textbf{0.7938}	&0.5645	&0.4366	&0.4740	&0.6562	&300\\
  shuttle	&\textbf{0.9971}	&0.7825	&0.8364	&0.4618	&0.8584	&1500\\
  page-blocks	&\textbf{0.9321} &0.7888	&0.9127	&0.4034	&0.8977	&300\\\bottomrule
\end{tabular}
\end{table*}

\begin{table*}[!h]
\footnotesize
\centering
\caption{ Time overhead on datasets (Unit: sec)}
\begin{tabular}{cccccccccccccccccccc}
  \bottomrule
  \multicolumn{1}{c}{\multirow{2}*{}}  &\multicolumn{1}{c}{\multirow{2}*{ECBE}} &\multicolumn{4}{c}{\multirow{1}*{OS-ELM}}\\
  \cline{3-6}
  &&radbas &sigmoid &sine &hardlim\\\midrule
  waveform	&144.6957	&74.5529	&72.5873	&\textbf{71.9789}	&72.6185 \\
  Hyperplane &\textbf{68.7527}	&72.9289	&74.4905	&773.6481	&73.6169\\
  LED	&\textbf{24.1768}	&73.3113 &72.6981	&73.9757	&72.5395\\
  sensor\_reading\_24	&0.8221	&0.3586	&\textbf{0.2652}	&0.3588	&0.3486\\
  shuttle	&\textbf{7.7122}	&99.4194	&24.1985	&24.1958	&24.1802\\
  page-blocks &\textbf{0.8741}	&0.6552	&0.2964	&0.2187	&0.3987\\\bottomrule
\end{tabular}
\end{table*}
From Table 7, it can be seen, for the accuracy, ECBE is better than OS-ELM on 4 datasets; on the \emph{waveform} and \emph{Hyperplane} datasets, when the activation function of OS-ELM is \emph{sigmoid}, OS-ELM is better than ECBE, but in fact, the advantage is very weak, after calculating, we can know the accuracies of OS-ELM are only more 0.0124 and 0.0194 than that of ECBE. In the aspect of time overhead, ECBE is better than OS-ELM on 4 datasets and OS-ELM is only better than ECBE on 2 datasets. If we synthetically consider the time overhead and accuracy of the two algorithms, on the \emph{sensor\_reading\_24}, although ECBE costs more time, the accuracy of ECBE is far higher than that of OS-ELM, so ECBE is still better than OS-ELM. From the analyses, it is obvious that ECBE is the best of all.

In order to verify the ability of ECBE detecting the concept drift, we select \emph{Hyperplane}, \emph{SEA}, \emph{waveform} and \emph{RBF} as experimental datasets which are produced by MOA. In consideration of concept drift in the datasets unknown to us, so we reprocess \emph{Hyperplane}, \emph{SEA} and \emph{waveform}, and make the three dataset contain 5 concept drift in which the label changes form 1 to 2 or reverse direction changing. The size of the four datasets are 3000, 3000, 3000 and 50000 respectively. The results are shown in Table 9 and Fig.8.

\begin{table*}
  \footnotesize
  \centering
  \caption{The number of concept drift detected by ECBE}
  \begin{tabular}{cccccccccccc}
  \bottomrule
  Dataset	&Hyperplane     &SEA	  &waveform	    &RBF  \\\midrule
  number	&6	            &4        &6	        &7  \\
  accuracy  &0.6420	        &0.7460	  &0.746	    &0.2002  \\
  winsize   &100            &100      &100          &1650\\\bottomrule
\end{tabular}
\end{table*}

\begin{figure*}
  \centering
  \includegraphics[width=10cm,height=6cm]{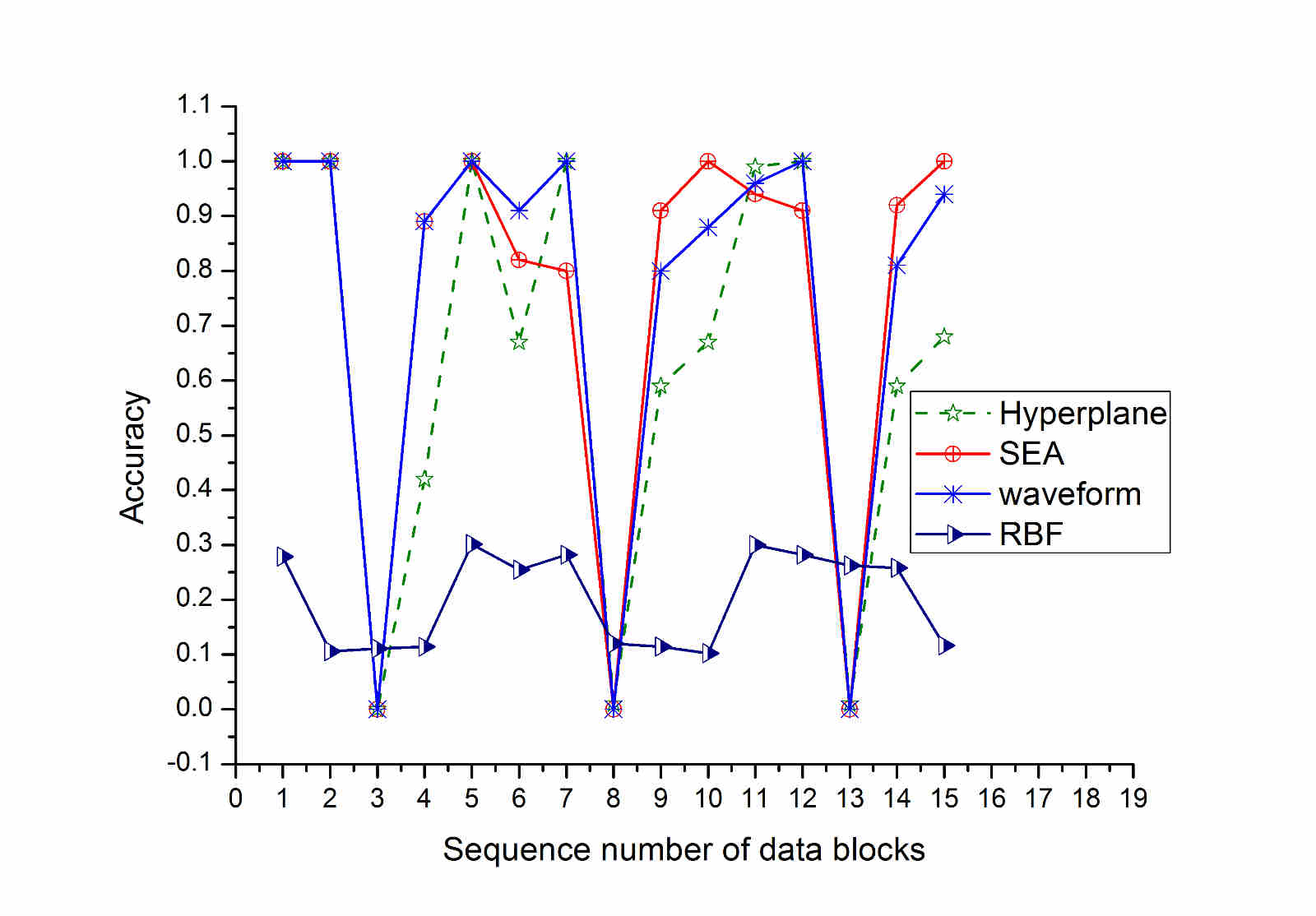}\\
  \caption{Test result with different datasets of ECBE}
\end{figure*}

From Fig.8, we can see when concept drift appearing, the accuracy of ECBE will degrade rapidly, and then ECBE will delete classifiers with weak performance, so the accuracy will be restored to the previous high level. In the experiment, ECBE detects concept drift for 15 times; from Table 9, we can know, on the Hyperplane and waveform datasets, the algorithm has detected all concept drifts, but producing a error alert; on the SEA dataset, no error alert has produced, but a real changing is omitted; RBF is a gradual concept drift dataset which is generated by computer and the number of concept drift in the datset is unknown; we use RBF to test the effectiveness of ECBE handling gradual concept drift; from the result on RBF, it is obvious that ECBE can detect gradual concept drift. So in summary, we can include that although there are some deviations for concept drift detection in the algorithm, the experimental data indicates the mechanism of ECBE detecting concept drift is effective and ECBE can cope with different types of concept drift.

\section{Conclusion}

In this paper, to solve the classification problem of the data streams, an ensemble classification algorithm based on information entropy was proposed. The algorithm is on the basis of the ensemble classification method, and utilizes the change of entropies before and after classification to detect concept drift. The experimental results show that ECBE is effective. But it is obvious that the ECBE algorithm is only suitable for the single labeled data, so how to apply the algorithm to the classification of multi-label data is the focus of our future research.

\section*{Acknowledgments}
This work was supported by National Natural Science Fund of China (Nos.U1435212, 61432011, 61202018, 61303008),National Key Basic Research and Development Program of China (973) (No.2013CB329404). The authors are grateful to the editor and the anonymous reviewers for constructive comments that helped to improve the quality and presentation of this paper.

\section*{References}
\scriptsize{
\my[1] J. Gama, R. Sebasti, P.P. Rodrigues, Issues in evaluation of stream learning algorithms,  Proceedings of the 15th ACM SIGKDD international conference on Knowledge discovery and data mining, (ACM, Paris, France, 2009), pp. 329-338.

\my[2] L.L. Minku, X. Yao, DDD: A New Ensemble Approach for Dealing with Concept Drift, IEEE Transactions on Knowledge \& Data Engineering, 24 (2011) 619-633.

\my [3] A. Bifet, G. Holmes, B. Pfahringer, R. Kirkby, Gavald, R. , New ensemble methods for evolving data streams,  ACM SIGKDD International Conference on Knowledge Discovery \& Data Mining (ACM 2009), pp. 139--148.

\my [4] Z Ouyang, M Zhou, et al. Mining concept-drifting and noisy data streams using ensemble classifiers. Artificial Intelligence and Computational Intelligence, 2009. AICI'09. International Conference on. Vol. 4. IEEE, 2009.

\my [5] X. Wu, P. Li, X. Hu, Learning from concept drifting data streams with unlabeled data, Neurocomputing, 92 (2012) 145-155

\my [6] D. Liu, Y. Wu, H. Jiang, FP-ELM: An online sequential learning algorithm for dealing with concept drift, Neurocomputing, 207 (2016) 322-334.

\my [7] P. Jedrzejowicz, I. Czarnowski, R.J. Howlett, L.C. Jain, I. Czarnowski, P. J drzejowicz, Ensemble Classifier for Mining Data Streams, Procedia Computer Science, 35 (2014) 397-406.

\my [8] P. Li, X. Wu, X. Hu, H. Wang, Learning concept-drifting data streams with random ensemble decision trees, Neurocomputing, 166 (2015) 68-83.

\my [9] R. Elwell, R. Polikar, Incremental Learning of Concept Drift in Nonstationary Environments, IEEE Transactions on Neural Networks, 22 (2011) 1517-1531.

\my [10] J. Rushing, S. Graves, E. Criswell, A. Lin, A Coverage Based Ensemble Algorithm (CBEA) for Streaming Data,  IEEE International Conference on TOOLS with Artificial Intelligence (2004), pp. 106-112.

\my [11] H. Murilo Gomes, F. Enembreck, SAE: Social Adaptive Ensemble classifier for data streams,  Computational Intelligence and Data Mining (2013), pp. 199-206.

\my [12] D. Brzezinski, J. Stefanowski, Combining block-based and online methods in learning ensembles from concept drifting data streams, Information Sciences An International Journal, 265 (2014) 50-67.

\my [13] D.M. Farid, L. Zhang, A. Hossain, C.M. Rahman, R. Strachan, G. Sexton, K. Dahal, An adaptive ensemble classifier for mining concept drifting data streams, Expert Systems with Applications, 40 (2013) 5895-5906.

\my [14] N.Y. Liang, G.B. Huang, P. Saratchandran, N. Sundararajan, A Fast and Accurate Online Sequential Learning Algorithm for Feedforward Networks, IEEE Transactions on Neural Networks, 17 (2006) 1411-1423.

\my [15] J.S. Lim, S. Lee, H.S. Pang, Low complexity adaptive forgetting factor for online sequential extreme learning machine (OS-ELM) for application to nonstationary system estimations, Neural Computing \& Applications, 22 (2013) 569-576.

\my [16] V. Kumar, P. Gaur, A.P. Mittal, Trajectory control of DC servo using OS-ELM based controller,  Power India Conference, 2012 IEEE Fifth (2012), pp. 1-5.

\my [17] Z. Yang, Q. Wu, C. Leung, C. Miao, OS-ELM Based Emotion Recognition for Empathetic Elderly Companion (Springer International Publishing, (2015).

\my [18] G.B. Huang, H. Zhou, X. Ding, R. Zhang, Extreme learning machine for regression and multiclass classification, IEEE Transactions on Systems Man \& Cybernetics Part B Cybernetics A Publication of the IEEE Systems Man \& Cybernetics Society, 42 (2012) 513-529.

\my [19] G.B. Huang, Q.Y. Zhu, C.K. Siew, Extreme learning machine: Theory and applications, Neurocomputing, 70 (2006) 489-501.

\my [20] Gu, Yang, et al., TOSELM: timeliness online sequential extreme learning machine. Neurocomputing 128 (2014): 119-127.

\my [21] Z. Ma, G. Luo, D. Huang, Short term traffic flow prediction based on on-line sequential extreme learning machine,  Eighth International Conference on Advanced Computational Intelligence (2016).

\my [22] P. Escandell-Montero, D. Lorente, J.M. Mart¨ªnez-Mart¨ªnez, E. Soria-Olivas, J. Vila-Frances, J.D. Mart¨ªn-Guerrero, Online fitted policy iteration based on extreme learning machines, Knowledge-Based Systems, 100 (2016) 200-211.

\my [23] H. Ryang, U. Yun, High utility pattern mining over data streams with sliding window technique, Expert Systems with Applications, 57 (2016) 214-231.

\my [24] H. Becker, M. Arias, Real-time ranking with concept drift using expert advice,  ACM SIGKDD International Conference on Knowledge Discovery and Data Mining (2007), pp. 86-94.

\my [25] J. Gama, P. Medas, P.P. Rodrigues, Learning decision trees from dynamic data streams,  ACM Symposium on Applied Computing (2005), pp. 685-688.

\my [26] D. Brzezinski, J. Stefanowski, Reacting to different types of concept drift: the Accuracy Updated Ensemble algorithm, IEEE Transactions on Neural Networks \& Learning Systems, 25 (2014) 81 - 94.

\my [27] H. Wang, W. Fan, P.S. Yu, J. Han, Mining concept-drifting data streams using ensemble classifiers, Kdd, (2003) 226-235.

\my [28] Z. Ouyang, M. Zhou, T. Wang, Q. Wu, Mining Concept-Drifting and Noisy Data Streams Using Ensemble Classifiers,  International Conference on Artificial Intelligence and Computational Intelligence (2009), pp. 360-364.

\my [29] Q. Wei, Z. Yang, Z. Junping, W. Yong, Mining multi-label concept-drifting data streams using ensemble classifiers,  International Conference on Fuzzy Systems \& Knowledge Discovery (2009), pp. 789.

\my [30] S. Ramamurthy, R. Bhatnagar, Tracking Recurrent Concept Drift in Streaming Data Using Ensemble Classifiers,  International Conference on Machine Learning and Applications (2007), pp. 404-409.

\my [31] C.E. Shannon, A mathematical theory of communication (McGraw-Hill, 1974).

\my [32] L. Rutkowski, L. Pietruczuk, P. Duda, M. Jaworski, Decision Trees for Mining Data Streams Based on the McDiarmid's Bound, Knowledge \& Data Engineering IEEE Transactions on, 25 (2013) 1272-1279.

\my [33] P. Domingos, G. Hulten, Mining high-speed data streams,  ACM SIGKDD International Conference on Knowledge Discovery and Data Mining (2000), pp. 71-80.

\my [34] W.N. Street, A streaming ensemble algorithm (SEA) for large-scale classification,  ACM SIGKDD International Conference on Knowledge Discovery \& Data Mining (2001), pp. 1152-1165.

\my [35] J.Z. Kolter, M.A. Maloof, Using additive expert ensembles to cope with concept drift,  International Conference, Bonn, Germany, August (2005), pp. 449--456.

\my [36] Y. Zhang, X. Jin, An automatic construction and organization strategy for ensemble learning on data streams, SIGMOD Rec., 35 (2006) 28-33.

\my [37] A. Bifet, G. Holmes, R. Kirkby, B. Pfahringer, MOA: Massive Online Analysis, Journal of Machine Learning Research, 11 (2010) 1601-1604.

\my [38] Xie Z, Tie Y, Guan L. A new audiovisual emotion recognition system using entropy-estimation-based multimodal information fusion, IEEE International Symposium on Circuits and Systems. IEEE, 2015:726-729.

\my [39] Zhang X, Mei C, Chen D, et al., Feature selection in mixed data: A method using a novel fuzzy rough set-based information entropy, Pattern Recognition, 2016, 56(1):1-15.

\my [40] Chen Y, Zhang Z, Zheng J, et al., Gene selection for tumor classification using neighborhood rough sets and entropy measures, Journal of Biomedical Informatics, 2017, 67:59-68.

\my [41] Y Li,D Li,S Wang,Y Zhai, Incremental entropy-based clustering on categorical data streams with concept drift, Knowledge-Based Systems, 2014, 59:33-47.

\my [42] Y Yao, L Feng, The Research on Massive and Dynamic Data Stream Classification Method, Dalian University of Technology, 2013. (in Chinese)

}

\end{document}